# Novel multiferroics with ferromagnetic phase induced in paraelectric antiferromagnets by electric field application

Maya D. Glinchuk[1], Eugene A. Eliseev[1], Yijia Gu[2], Long-Qing Chen[2], Venkatraman Gopalan[2*]

and Anna N. Morozovska[1,3†]

[1] Institute for Problems of Materials Science, NAS of Ukraine,

Krjijanovskogo 3, 03142 Kiev, Ukraine

[2] Department of Materials Science and Engineering, Pennsylvania State University, University

Park, PA 16802, USA [3] Institute of Physics, NAS of Ukraine, 46, pr. Nauki, 03028 Kiev,

Ukraine

**Abstract**

The phase diagram of a quantum paraelectric antiferromagnet $EuTiO_3$ under an external electric field was calculated using Landau-Ginzburg-Devonshire theory. It was shown that the application of an external electric field $E$ leads to the appearance of a ferromagnetic phase due to the magnetoelectric coupling. In particular, electric field application decreases the transition temperature $T_{AFM}$ to antiferromagnetic (AFM) phase and induces ferromagnetic (FM) phase, so that at some $E$ field larger than the critical field ($E_{cr}$), $T_{FM}$ becomes higher than $T_{AFM}$ and the FM phase appears. Note that $E_{cr}$ increases and magnetization decreases as the temperature increases. The value of the critical field $E_{cr} = 0.40 \times 10^6$ V/cm we calculated appeared close to the value $E_{cr} = 0.5 \times 10^6$ V/cm obtained recently by Ryan et al. with the help of density functional theory for $EuTiO_3$ film under a compressive strain produced by substrate. At the fields $E \geq 0.83 \times 10^6$ V/cm, AFM disappears for all considered temperatures and so FM becomes the only stable magnetic phase.

We find that ferromagnetic phase can be induced by an $E$-field in other paraelectric antiferromagnet oxides with a positive AFM-type magnetoelectric (ME) coupling coefficient and negative FM-type ME coupling coefficient. In particular, the critical E-field was estimated for another paraelectric antiferromagnet $Sr_{0.7}Ba_{0.3}MnO_3$ as $0.2 \times 10^5$ V/cm at 0 K.

Analysis of the dependence of magnetization and antimagnetization on the external electric field and the polarization induced by the field, which yields the magnetoelectric coupling, is reported. The results show the possibility to control multiferroicity, including the FM and AFM phases, with help of an electric field application.

[*] Corresponding author1: vxg8@psu.edu
[†] Corresponding author2: anna.n.morozovska@gmail.com

# 1. Introduction

The search for new multiferroic materials with large magnetoelectric (ME) coupling is very interesting for fundamental studies and is important for applications. For example, based on the magnetic field control of the dielectric properties of the material, information can be recorded by an electric field and non-destructively read by a magnetic field [1, 2, 3, 4].

In the last few years, $EuTiO_3$ has been extensively studied as a basis for discovering new multiferroics. The bulk quantum paraelectric $EuTiO_3$ is a low temperature antiferromagnet [1, 2]. It exhibits an antiferrodistorsive (AFD) transition at 281 K [5, 6, 7, 8, 9] and is paraelectric at all temperatures. The strained $EuTiO_3$ films, surprisingly, become a strong ferroelectric ferromagnet under epitaxial tensile strains exceeding 1 % [10, 11]. Recently a lot of attention has been paid to the impact of the structural AFD order parameter (oxygen octahedron static rotations [12]) on phase diagrams, structural, polar and magnetic properties of $EuTiO_3$ and its solid solution with another quantum paraelectric $SrTiO_3$, namely $Eu_xSr_{1-x}TiO_3$ [13]. In particular, a complex interplay between the AFD order parameter and electric polarization in tensile strained $Eu_xSr_{1-x}TiO_3$ thin films leads to the appearance of low-symmetry monoclinic phase with in-plane ferroelectric polarization [14]. Another important possibility in $Eu_xSr_{1-x}TiO_3$ solid solution is to control the appearance of FM phase by changing the concentration of Sr ions. Since the dilution of magnetic Eu ions by nonmagnetic Sr ions might change the type of Eu ions magnetic order because of different percolation thresholds for ferromagnetic (FM) ($x_{cr}^F \approx 0.24$) and antiferromagnetic (AFM) ($x_{cr}^A \approx 0.48$) order, a FM phase may become stable at some finite concentration of Sr ions [15].

Several years ago electric field $E$ control of ferromagnetism was a hot topic for the scientists around the world (see e.g. [16, 17] and references therein). For the case of semiconductors with a hole-induced ferromagnetism, the influence of $E$-field on the carrier properties is considered [18, 19, 20]. Recenlty, Ryan et al. [17] considered the possibility of a reversible control of magnetic interactions in $EuTiO_3$ by applying $E$-field. Because of Ti displacement from its central position under the $E$-field, changes in the spatial overlap between the electronic orbitals of the ions, and thus the magnetic exchange coupling is expected. In particular, the density functional theory calculation shows that the competition between FM and AFM interactions is resolved in favour of FM for paraelectric $EuTiO_3$ film on compressive substrate, when applied $E$-field exceeds a critical value estimated as $E_{cr} = 0.5 \times 10^6$ V/cm. It is obvious that the mechanism proposed in Ref. [17] is based on the magnetoelectric coupling.

These facts motivate us to perform analytical calculations of the influence of the $E$-field on the $EuTiO_3$ phase diagram in the framework of Landau–Ginzburg–Devonshire (LGD) theory [21, 22, 23, 24, 25, 26]. Below we consider the magnetoelectric coupling [27] characteristic for EuTiO$_3$ as the main mechanism of *E*-field influence on the phase diagram. We analyze and compare the magnetization and antimagnetization dependence on the polarization induced by an external *E*-field and on the magnetoelectric coupling. Our analytical results have shown the possibility to control multiferroicity, including the FM and AFM phases with applied electric field in different paraelectric antiferromagnets under certain conditions imposed on the ME coupling coefficients.

## 2. Electric field induced ferromagnetism in bulk EuTiO$_3$

Let us study the possibility of electric field induced ferromagnetism in bulk EuTiO$_3$ using LGD theory. For the considered case, LGD approach is based on the phase stability analysis of thermodynamic potential (free energy) that is a series expansion to various powers of the order parameters (polarization and magnetization). The magnetization and polarization-dependent part of the corresponding free energy is [15, 28]:

$$G_M = \int_V d^3r \left( \begin{array}{c} \dfrac{\alpha_P}{2} P_3^2 + \dfrac{\beta_P}{4} P_3^4 - E_3 P_3 + \dfrac{\alpha_M}{2} M^2 + \dfrac{\alpha_L}{2} L^2 + \dfrac{\beta_M}{4} M^4 + \dfrac{\beta_L}{4} L^4 \\ + \dfrac{\lambda}{2} L^2 M^2 + \dfrac{P_3^2}{2} \left( \eta_{FM} M^2 + \eta_{AFM} L^2 \right) \end{array} \right) \quad (1)$$

Here $P_3$ is ferroelectric polarization component, $E_3$ is external electric field component, $M^2 = M_1^2 + M_2^2 + M_3^2$ is ferromagnetic magnetization square and $L^2 = L_1^2 + L_2^2 + L_3^2$ is antiferromagnetic magnetization square correspondingly. The last two terms represent biquadratic ME coupling between order parameters.

Expansion coefficient $\alpha_P$ depends on the absolute temperature *T* in accordance with Barrett law, namely $\alpha_P(T) = \alpha_T^{(P)} \left( T_q^{(P)}/2 \right) \left( \coth\left( T_q^{(P)}/2T \right) - \coth\left( T_q^{(P)}/2T_c^{(P)} \right) \right)$. Here $\alpha_T^{(P)}$ is constant, temperatures $T_q^{(P)}$ is the so-called quantum vibration temperature related with polar soft modes, $T_c^{(P)}$ is the "effective" Curie temperature corresponding to the polar modes in bulk EuTiO$_3$. Coefficient $\beta_P$ is regarded as temperature independent [15].

Expansion coefficient $\alpha_M$ depends on the temperature in accordance with the Curie law, namely $\alpha_M(T) = \alpha_C(T - T_C)$, where $T_C$ is the ferromagnetic Curie temperature. Note that the dependence determines the experimentally observed inverse magnetic susceptibility in paramagnetic phase of EuTiO$_3$. The temperature dependence of the expansion coefficient $\alpha_L$ is $\alpha_L(T) = \alpha_N(T - T_N)$, where $T_N$ is the Neel temperature for bulk EuTiO$_3$. For equivalent permutated magnetic Eu ions with antiparallel spin ordering, it can be assumed that $\alpha_C \approx \alpha_N$.

The LM-coupling coefficient $\lambda$ should be positive, because only the positive coupling term $\lambda L^2 M^2/2$ prevents the appearance of ferromagnetic (as well as ferrimagnetic) phases at low temperatures $T < T_C$ under the condition of $\sqrt{\beta_M \beta_L} < \lambda$ regarded valid hereafter [17]. Coefficients $\beta_L$ and $\beta_M$ are regarded as positive and temperature independent.

Biquadratic ME coupling contribution is $(\eta_{FM} M^2 + \eta_{AFM} L^2) P_3^2/2$. Following Lee *et al.* [11] we assume that ME coupling coefficients of FM and AFM are equal and positive, i.e. $\eta_{AFM} \approx -\eta_{FM} > 0$ for numerical calculations, as anticipated for equivalent magnetic Eu ions with antiparallel spin ordering in a bulk EuTiO$_3$.

Considering the case of incipient ferroelectric and in order to obtain analytical results, one could suppose a linear dependence of polarization on applied electric field

$$P_3 \approx \chi E_3 \tag{2}$$

Here we introduce linear dielectric susceptibility $\chi$ as

$$\chi = \frac{1}{\alpha_P + \eta_{FM} M^2 + \eta_{AFM} L^2} \tag{3}$$

Equations of state for the absolute value of the magnetization $M$, and the antimagentization $L$ can be obtained from the minimization of the free energy (1). They are $(\alpha_M + \eta_{FM} P_3^2) M + \beta_M M^3 + \lambda L^2 M = 0$ and $(\alpha_L + \eta_{AFM} P_3^2) L + \beta_L L^3 + \lambda L M^2 = 0$. The formal solution of these equations contains the possible $E$-field induced phase transition, namely the appearance of the mixed ferromagnetic phase with order parameters:

$$M = \sqrt{\frac{\alpha_L \lambda - \alpha_M \beta_L + (\lambda \eta_{AFM} - \beta_L \eta_{FM}) P^2}{\beta_M \beta_L - \lambda^2}}, \tag{4a}$$

$$L = \sqrt{\frac{\alpha_M \lambda - \alpha_L \beta_M + (\lambda \eta_{FM} - \beta_M \eta_{AFM}) P^2}{\beta_M \beta_L - \lambda^2}} \tag{4b}$$

The critical values of polarization could be found by substituting into the equations either $M=0$ or $L=0$, i.e.:

$$P_{cr}\big|_{M=0} = \sqrt{\frac{\alpha_M \beta_L - \alpha_L \lambda}{\lambda \eta_{AFM} - \beta_L \eta_{FM}}}, \quad P_{cr}\big|_{L=0} = \sqrt{\frac{\alpha_L \beta_M - \alpha_M \lambda}{\lambda \eta_{FM} - \beta_M \eta_{AFM}}}. \tag{5}$$

Expressions (5) correspond to the lower and upper critical fields respectively:

$$E_{cr}\big|_{M=0} = \frac{1}{\chi}\sqrt{\frac{\alpha_M \beta_L - \alpha_L \lambda}{\lambda \eta_{AFM} - \beta_L \eta_{FM}}}, \quad E_{cr}\big|_{L=0} = \frac{1}{\chi}\sqrt{\frac{\alpha_L \beta_M - \alpha_M \lambda}{\lambda \eta_{FM} - \beta_M \eta_{AFM}}} \tag{6}$$

Note, that LM-coupling constant λ, $β_M$ and $β_L$ are positive as required for the stability of free energy (1). Using the conditions in the expressions (6), the conditions $η_{FM} < 0$ and $η_{AFM} > 0$, $α_M(T) > 0$ and $α_L(T) < 0$, are sufficient for the absolute stability of the FM phase at applied electric fields greater than the critical field $E_{cr}|_{M=0}$, and with arbitrary positive values of λ and $β_{L,M}$. Note, that under the typical condition of small positive LM-coupling constant λ, one immediately obtains from Eqs.(6), simpler equations $E_{cr}|_{M=0} \approx \chi^{-1}\sqrt{-α_M/η_{FM}}$ and $E_{cr}|_{L=0} \approx \chi^{-1}\sqrt{-α_L/η_{AFM}}$ that are useful for estimations.

Using EuTiO$_3$ parameters listed in the **Table 1**, one could see that the condition $E > E_{cr}|_{M=0}$ becomes valid for electric fields higher than 480 kV/cm at 0 K. The value is in reasonable agreement with DFT simulations performed by Ryan et al. [17].

**Table 1**. Polarization and magnetic parts of the free energy for EuTiO$_3$

| Parameter | SI units | Value |
|---|---|---|
| coefficient $α_T^{(P)}$ | $10^6$ m/(F K) | 1.95 |
| Effective Curie temperature $T_c^{(P)}$ | K | -133.5 |
| Characteristic temperature $T_q^{(P)}$ | K | 230 |
| LGD-coefficient $β_P$ | $10^9$ m$^5$/(C$^2$F) | 1.6 |
| LGD-coefficient $α_C \approx α_N$ * | Henri/(m·K) | $2π·10^{-6}$ |
| LGD-coefficient $β_M$ | J m/A$^4$ | $0.8×10^{-16}$ |
| LGD-coefficient $β_L$ | J m/A$^4$ | $1.33×10^{-16}$ |
| LGD-coefficient λ | J m/A$^4$ | $1.0×10^{-16}$ |
| AFM Neel temperature $T_N$ | K | 5.5 |
| FM Curie temperature $T_C$ | K | 3.5±0.3 |
| ME coupling coefficient $η_{AFM}$ | J m$^3$/(C$^2$ A$^2$) | $8×10^{-5}$ |
| ME coupling coefficient $η_{FM}$ | J m$^3$/(C$^2$ A$^2$) | $-8×10^{-5}$ |

*) the equality comes from the equivalence of the magnetic sub lattices in EuTiO$_3$

The complex behaviour of $M$ and $L$ induced by $E_3$ can be explained by the phase diagram of bulk EuTiO$_3$ in the coordinates of temperature and external electric field, as shown in **Figure 1a**. Note that all the magnetic phases also possess an AFD ordering and the influence is included in the renormalization of the LGD-expansion coefficients listed in **Table 1**. One can see from the diagram that the FM phase stability region starts at electric fields greater than 0.5 MV/cm at 0 K, and converges to 0.83 MV/cm at 4 K. Paramagnetic (PM) phase is stable at temperatures greater than 5 K, while its boundary with AFM phase slightly shifts to the lower temperatures as the electric field increases. Triangular-like region of the AFM phase exists between the FM and PM phases at temperatures lower that 5 K and electric field less than 0.83 MV/cm. A rather thin wedge-like region of the ferrimagnetic (FI) phase exists between the FM and AFM phase at temperatures less than 3 K and for fields between 0.4 MV/cm to 0.7 MV/cm. At a field of $E_{cr} \geq 0.83$ MV/cm, the AFM phase disappears at all considered temperatures and so the true FM phase becomes the only absolutely stable magnetic phase. The phase diagram proves that an electric field higher than $E_{cr}$ transforms the bulk EuTiO$_3$ into a true and relatively strong FM state at temperatures lower than 5 K. The result opens up the possibility to control bulk EuTiO$_3$ between different magnetic phases using external electric field. In particular, our calculations prove that it becomes possible to control the multiferroicity, including the content of FM and AFM phases, with the help of external electric fields. Note, that **Figure 1a** addresses the question of which phase (FM, FI, AFM or PM) is absolutely stable at a given temperature and electric field.

**Figure 1b** illustrates the temperature dependences of the magnetization $M$ (solid curves) and anti-magnetization $L$ (dashed curves) for different values of external electric field. Mostly there are regions where both $M$ and $L$ coexist, indicating the presence of electric field – induced FI phase that microscopically could be realized as canted AFM phase as anticipated for equivalent magnetic Eu ions with antiparallel spin ordering at zero field. For realistic values of applied field, the temperature range of the FM phase is below 5 K. As one can see from the **Fig. 1b** the temperature interval for the existence of magnetization $M$ increases with applied electric field. For example, the transition to FM phase occurs at 1 K for the field $E_3 = 0.5$ MV/cm and at 5 K for $E_3 = 0.9$ MV/cm. Note that the FM transition is of the second order for $E_3 = 0.9$ MV/cm, but it is more close to the first order at smaller $E_3$. The magnetization value, which is ~0.5 A/m at the low temperatures is higher than the corresponding antimagnetization. Antimagnetization, $L$, completely disappears for $E_3 = 0.83$ MV/cm, and only magnetization exists at this field. Antimagnetization $L$ disappears at about 5 K and electric field $E_{crL}=0.65$ MV/cm (the second order phase transition). The points of the second order phase transition depend on the $E_3$ value rather weakly. In contrast to magnetization that appears at low temperatures and then only increases as the temperature decreases down to 0 K, the antimagnetization $L$ does not exist at low temperatures $T < 0.5$ K; it typically appears at the temperature range with no magnetization (compare solid and dashed curves in the **Fig. 1b**). The higher the electric field, the smaller the temperature region of nonzero antimagnetization: For example, it exists from 0.5 K-5 K at $E_3 = 0.5$ MV/cm and from 3.5 K-4.5 K at $E_3 = 0.9$ MV/cm.

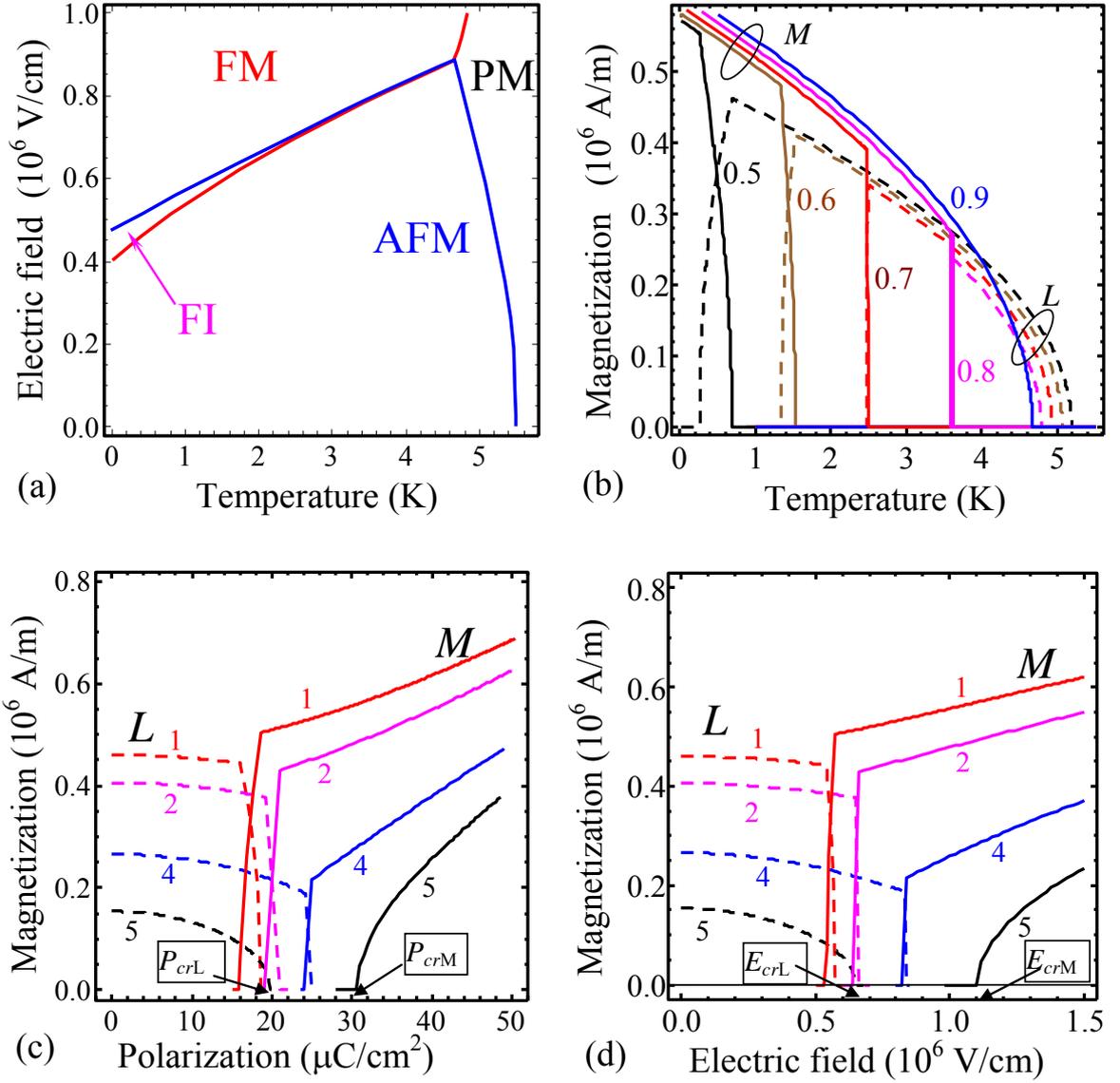

**Figure 1**. (a) Phase diagram of bulk $EuTiO_3$ in coordinates temperature – external electric field. PM – paramagnetic phase, FI – ferrimagnetic phase, FM – ferromagnetic phase, AFM – antiferromagnetic phase. (b) Temperature dependences of the magnetization $M$ (solid curves) and anti-magnetization $L$ (dashed curves) for different values of external electric field $E_3$=0.5, 0.6, 0.7, 0.8 and 0.9 MV/cm (numbers near the curves). (c,d) Magnetization $M$ (solid curves) and anti-magnetization $L$ (dashed curves) as a function of polarization (d) induced by an external electric filed (c) at different temperatures (values in Kelvins are shown near the curves).

**Figures 1c** and **1d** illustrate $M$ (solid curves) and anti-magnetization $L$ (dashed curves) as a function of polarization induced by an external electric field and as a function of electric field itself at different temperatures from 1 – 5 K. One can see that at electric fields less than the critical value, only AFM magnetization exists. For the electric fields greater than the critical value, a ferromagnetic magnetization occurs and increases as the strength of the electric field (or polarization) increases. An unusual cross-over from the first-order phase transition (corresponding to the FM magnetization appearance), to a second order transition appears with an increase in temperature. The decrease in antimagnetization for electric fields greater than the critical value follows the first order transition. The critical field value increases and the "gap" between the AFM and FM states shrinks with temperature increase (compare the curves calculated for 1 K with the ones for 4 K). At temperatures of 1 – 4 K, the thin region of $M$ and $L$ coexistence, i.e. FI phase, is seen. At 5 K, there is a pronounced gap between AFM and FM states.

Note that the critical value of polarization and E-field depend on the temperature as one can see from the examples shown in **Figs. 1c** and **1d**. The comparison of the x-axis in the **Figs 1c** and **1d** provides insight into the polarization values induced by electric field. Polarization below $P_{cr}$ (or sub-critical electric fields) cannot induce the ferromagnetism in bulk EuTiO$_3$. At the same time, a polarization value higher than the critical one induces FM with rather high $M$ values (up to 0.6 MA/m). The LGD approach makes it possible to calculate the corresponding phase diagram of the dependence of stable magnetic phases on the applied electric field and magnetoelectric characteristics.

### 3. Discussion and conclusion

The phase diagram proves that electric field higher than $E_{cr}$ transforms the bulk EuTiO$_3$ into a true and relatively strong ferromagnetic state at temperatures lower 5 K. Therefore, it can be shown that at fields $E$ greater than $E_{cr} = 0.83$ MV/cm and temperatures $T < 5$ K, the transition temperature to FM phase becomes higher than the transition temperature to AFM phase. Allowing for the value of transition temperature depends on the superexchange of Eu-Ti-Eu bond alignment and the degree of interatomic orbital overlap, the distortion induced by electric field could significantly alter the magnetic structure of the entire system. Such behavior follows from the **Fig. 1c** and **d**, where the dependence of magnetization and antimagnetization on polarization and electric field is presented for several temperatures. Also from the **Fig. 1c** and **d**, one can see that the magnetization increases as the polarization increases, while the antimagnetization decreases. Thus, the shift of Ti ions from central position induced by the electric field disrupts the long-range spin coherence of AFM order originated from the third Eu ion neighbours interaction, meanwhile the first and the second neighbouring Eu ions are ferromagnetically ordered in accordance with the results of Ryan et al [17].

Generally speaking, one can look for the fulfilment of expressions (6) in other paraelectric antiferromagnet oxides with ME coupling coefficients satisfying the conditions $\eta_{FM} < 0$ and $\eta_{AFM} > 0$, where the magnetization could be induced by an electric field $E > E_{cr}|_{M=0}$ (where $E_{cr}|_{M=0}$ is given by Eq.(6)), at some temperature range defined by the conditions $\alpha_M(T) > 0$ and $\alpha_L(T) < 0$. The search for such materials seems to be important both for understanding the mechanisms of ME coupling and for possible applications. The main problem is the restricted knowledge about ME coupling coefficients.

Let us discuss some cases when one can expect *E*-field induced magnetization. In particular such supposition can be made on the basis of data known for solid solutions $Sr_{1-x}Ba_xMnO_3$ [29] and $Sr_{1-x}Eu_xTiO_3$ [15]. Sakai et al [29] had shown the strong suppression of ferroelectricity observed at x≥0.4 and originated from $Mn^{+4}$ ions displacement upon the antiferromagnetic order. This gives the direct evidence that $\eta_{AFM} > 0$ and with respect to the above written expression for the critical field, $\eta_{FM} < 0$. The assumption about different signs of $\eta_{FM}$ and $\eta_{AFM}$ also agrees with Smolenskii and Chupis [30], as well as Katsufuji [1] and Lee et al [11]. In particular Smolenskii and Chupis and Katsufuji [1] stated that it is natural to consider that the dielectric constant is dominated by the pair correlation between the nearest spins, which phenomenologically leads to the ME term $\eta P^2(M^2 - L^2)$. Therefore we arrived at the conclusion about the fulfilment of conditions (6) and hence the possibility to induce magnetization by *E*-field at x<0.4, e.g. x=0.2, where we have a paraelectric antiferromagnet. Following **Table 2** we calculated the critical *E*-field for $Sr_{0.7}Ba_{0.3}MnO_3$ which is equal to $0.2 \cdot 10^5$ V/cm at 0 K.

Another solid solution $(Eu,Sr)TiO_3$ considered by us [15] based on $EuTiO_3$ could have its ME characteristics $\eta_{FM} < 0$ and $\eta_{AFM} > 0$ (see the **Table 1**). Since it is possible to have FM phase at x between percolation thresholds of FM and AFM phases (x=0.24 and x=0.48 respectively [15]), one has to look for *E*-field induced FM phase, e.g. at x<0.2.

Note that in Table 2, we did not include the already mentioned $E_{cr}=0.5\times10^6$ V/cm for $EuTiO_3$ film on the compressive substrate [17] and $Sr_xEu_{1-x}TiO_3$ paraelectric antiferromagnet for x<0.2 since we do not know the values for all necessary parameters. The search for the other paraelectric antiferromagnets, which satisfy the conditions $\eta_{FM} < 0$ and $\eta_{AFM} > 0$, is in progress.

The experimental confirmation of the theoretical prediction is extremely desirable. The observation of the ferromagnetic phase under an electric field or induced by polarization due to $Mn^{+4}$ or $Ti^{+4}$ ions shift will provide a direct evidence of magnetoelectric coupling mechanism based on the superexchange of Mn-O-Mn or Eu-Ti-Eu bonds proposed recently in Ref. [17, 29] based on first principle calculations.

**Table 2.** ME coupling coefficients and critical E-fields for some paraelectric antiferromagnets.

| Paraelectric antiferromagnet | Neel temperature $T_N$ (K) | ME coupling coefficients J m$^3$/(C$^2$ A$^2$) | Critical field at 0 K (V/cm) | Ref. |
|---|---|---|---|---|
| $Sr_{1-x}Ba_xMnO_3$ ($x = 0.3$) | 215 | $\eta_{AFM} = -\eta_{FM} = 4\times10^{-2}$ * | $0.2\ 10^5$ | [29, 31] |
| $EuTiO_3$ | 5.5 | $\eta_{AFM} = -\eta_{FM} = 8\times10^{-5}$ | $0.4\ 10^6$ | table 1 |
| $MnTiO_3$ | 65 | $\eta_{FM}^{\parallel}=0.47$, $\eta_{FM}^{\perp}=0.18$ $\eta_{AFM}$ - not found | Does not exist because $\eta_{FM} > 0$ | [32, 33] |

*the spin-phonon coupling in $Sr_{1-x}Ba_xMnO_3$ ($x =0.3$) is more than 500 times stronger than that for $EuTiO_3$.

**Acknowledgements**

Authors acknowledge the financial support via bilateral SFFR-NSF project, namely US National Science Foundation under NSF-DMR-1210588 and State Fund of Fundamental Research of Ukraine, grant UU48/002.